\documentclass{webofc}%
\usepackage[varg]{txfonts} 
\usepackage{amssymb}
\usepackage[varg]{txfonts}
\usepackage{amsmath}
\usepackage{amsfonts}
\usepackage{graphicx}%
\providecommand{\U}[1]{\protect\rule{.1in}{.1in}}
\begin{document}

\title{Conventional mesons below 2 GeV}
\author{Francesco Giacosa \inst{1}\fnsep\thanks{{francescogiacosa@gmail.com}} }


\institute{Institute of Physics, Jan Kochanowski University,
ul. Uniwersytecka 7, 25-406, Kielce, Poland
}

\abstract{ We briefly review the status of various conventional quark-antiquark mesons below 2 GeV and outline some open questions: the status of the strange-antistrange orbitally excited vector meson, the status of the nonet of axial-tensor mesons (chiral partners of the well known tensor mesons), and the isoscalar mixing angle in the pseudotensor sector, which can eventually represent a novel manifestation of the chiral anomaly. 
}

\maketitle

\section{Introduction}

\label{intro}

The study of exotic mesons is an important topic in modern hadronic physics,
which is in the centre of dedicated theoretical and experimental works, e.g Refs.
\cite{pdg,pelaezrev,maiani,mai,bes,clas,gluex,panda,jpac} and refs. therein.

Yet, a necessary condition toward such a study is a clear understanding of
the conventional mesons and baryons, e.g. Refs. \cite{isgur,fischer}. Namely,
only when the standard quark-antiquark and three-quark states are fully under
control in a certain energy region, the search for states that do not fit into
this conventional picture can be fruitful.

In this work we concentrate on conventional mesonic states below 2 GeV. In
particular, we present a mini-review of the achievements of a series of works
(by myself and collaborators) based on a quite long time span in which various
mesonic nonets with different quantum numbers have been studied in the
framework of hadronic effective models that employ either flavor or chiral
symmetry. In particular, for any given nonet or chiral multiplet, either a
flavor or a chiral invariant and purely mesonic model has been set to study
the masses and, most importantly, the decays of the corresponding states. In this way one
could test the goodness of the quark-antiquark assignment by comparing
theoretical data to experimental decay widths and branching ratios.

\section{Quark-antiquark conventional nonets}

In the non-relativistic notation, a quark-antiquark ($\bar{q}q$) nonet is classified by
$n^{2S+1}L_{J}$, where $n$ is the radial quantum number, and $S$, $L$ and $J$
are the spin, spacial, and total angular momenta, respectively. 
The relativistic notation is denoted as
$J^{PC}$, where $P=(-1)^{L+1}$ is parity and $C=(-1)^{L+S}$ the charge conjugation.

In Table 1 we report various $n=1$ mesonic nonets together with both
notations above. We also refer to the results obtained by our previous
works on the subject by using effective flavor/chiral mesonic models that study
masses and decays. 
Here, we concentrate on the main picture and do not write
down the Lagrangian(s), for which we refer to the quoted papers. Moreover, for
each nonet we shall also discuss the mixing of the nonstrange and the strange
components, since this information may be linked to nonperturbative physics
(chiral anomaly).

\bigskip

\begin{center}
Table 1: Summary of conventional mesons and related references based on
hadronic models.%

\begin{tabular}
[c]{||l||l||l||l||l||l||l||l||}\hline\hline
$n^{2S+1}L_{J}$ & $J^{PC}$ & Current &
\begin{tabular}
[c]{@{}l}%
$I=1$\\
$u\overline{d}$, $d\overline{u}$\\
$\frac{d\overline{d}-u\overline{u}}{\sqrt{2}}$%
\end{tabular}
&
\begin{tabular}
[c]{@{}l}%
$I=1/2$\\
$u\overline{s}$, $d\overline{s}$\\
$s\overline{d}$, $s\overline{u}$%
\end{tabular}
&
\begin{tabular}
[c]{@{}l}%
$I=0$\\
$\approx\frac{u\overline{u}+d\overline{d}}{\sqrt{2}}$%
\end{tabular}
&
\begin{tabular}
[c]{@{}l}%
$I=0$\\
$\approx s\overline{s}$%
\end{tabular}
& Refs.\\\hline\hline
$1^{1}S_{0}$ & $0^{-+}$ & $\bar{q}i\gamma^{5}q$ & $\pi$ & $K$ & $\eta(547)$ &
$\eta^{\prime}(958)$ & \cite{dick}\\\hline\hline
$1^{3}P_{0}$ & $0^{++}$ & $\bar{q}q$ & $a_{0}(1450)$ & $K_{0}^{\star}(1430)$ &
$f_{0}(1370)$ & $f_{0}(1500)$ & \cite{dick,janowski,chiralplb,longchiral}%
\\\hline\hline
$1^{3}S_{1}$ & $1^{--}$ & $\bar{q}\gamma^{\mu}q$ & $\rho(770)$ & $K^{\star
}(892)$ & $\omega(782)$ & $\phi(1020)$ & \cite{dick}\\\hline\hline
$1^{3}P_{1}$ & $1^{++}$ & $\bar{q}\gamma^{5}\gamma^{\mu}q$ & $a_{1}(1260)$ &
$K_{1A}$ & $f_{1}(1285)$ & $f_{1}^{\prime}(1420)$ & \cite{dick,divotgey}%
\\\hline\hline
$1^{1}P_{1}$ & $1^{+-}$ & $\bar{q}\gamma^{5}\partial^{\mu}q$ & $b_{1}(1235)$ &
$K_{1B}$ & $h_{1}(1170)$ & $h_{1}(1415)$ & \cite{divotgey,sammet}%
\\\hline\hline
$1^{3}D_{1}$ & $1^{--}$ & $\bar{q}\partial^{\mu}q$ & $\rho(1700)$ & $K^{\star
}(1680)$ & $\omega(1650)$ & $\phi(???)$ & \cite{piotrowska,sammet}%
\\\hline\hline
$1^{3}P_{2}$ & $2^{++}$ & $\bar{q}i\gamma^{\mu}\partial^{\nu}q$ &
$a_{2}(1320)$ & $K_{2}^{\star}(1430)$ & $f_{2}(1270)$ & $f_{2}^{\prime}(1525)$
& \cite{tensor,axialtensors}\\\hline\hline
$1^{3}D_{2}$ & $2^{--}$ & $\bar{q}i\gamma^{5}\gamma^{\mu}\partial^{\nu}q$ &
$\rho_{2}(???)$ & $K_{2}(1820)$ & $\omega_{2}(???)$ & $\phi_{2}(???)$ &
\cite{axialtensors}\\\hline\hline
$1^{1}D_{2}$ & $2^{-+}$ & $\bar{q}i\gamma^{5}\partial^{\mu}\partial^{\nu}q$ &
$\pi_{2}(1670)$ & $K_{2}(1770)$ & $\eta_{2}(1645)$ & $\eta_{2}(1870)$ &
\cite{pseudotensor,shastry}\\\hline\hline
$1^{3}D_{3}$ & $3^{--}$ & $\bar{q}\gamma^{\mu}\partial^{\nu}\partial^{\rho}q$
& $\rho_{3}(1690)$ & $K_{3}^{\star}(1780)$ & $\omega_{3}(1670)$ & $\phi
_{3}(1850)$ & \cite{j3}\\\hline\hline
\end{tabular}

\end{center}

It is useful to comment one by one the entries of the Table 1 above:

\begin{itemize}
\item The first entry refers to the well-known and established pseudoscalar
mesons $\{\pi$, $K,\eta(547),\eta^{\prime}(958)\}$ with $J^{PC}=0^{-+}$. These
states are particularly important in low-energy QCD since they correspond to
the (quasi-)Goldstone bosons emerging upon spontaneous symmetry breaking of chiral symmetry.
In chiral mesonic models, they appear in (extensions of) the Mexican hat
potential \cite{beyond} and are also the starting point of chiral perturbation theory, e.g. Ref. \cite{chpt}. 
Due to the axial (or chiral) anomaly \cite{feldmann,thooft}, the
two isoscalar fields are given by the mixing:
\begin{equation}
\left(
\begin{array}
[c]{c}%
\eta(547)\\
\eta^{\prime}\equiv\eta(958)
\end{array}
\right)  =\left(
\begin{array}
[c]{cc}%
\cos\beta_{P} & \sin\beta_{P}\\
-\sin\beta_{P} & \cos\beta_{P}%
\end{array}
\right)  \left(
\begin{array}
[c]{c}%
\eta_{N}\equiv\sqrt{\frac{1}{2}}\,(\bar{u}u+\bar{d}d)\\
\eta_{S}\equiv\bar{s}s
\end{array}
\right)  \text{ ,}%
\end{equation}
with a large mixing angle $\beta_{P}=-43.4^{\circ}$ \cite{kloe2}. In other
words, the large mixing can be also understood by the fact that pseudoscalar mesons belong to a so-called \textquotedblleft
heterochiral\textquotedblright\ multiplet \cite{pisarski}, for which a chirally invariant (but axial breaking) term can be easily written down.

\item The second entry refers to their chiral partners of the pseudoscalar mesons, the since long time debated $\bar{q}q$ scalar mesons with
$J^{PC}=0^{++}$, see also e.g. Refs.
\cite{rodas,klemptnew,sarantsev,amsler}. Here, the assignment is not
yet conclusive, even if their placement above 1 GeV seems quite natural for $P$-wave
states. In addition, the scalar glueball, to be predominantly identified with
$f_{0}(1710)$  \cite{weingarten,gui,janowski} is likely to mix with these states. In turn, the scalar states
below 1 GeV need to be interpreted as four-quark objects \cite{pelaezrev}.

\item In the third entry, the well-known vector mesons \{$\rho(770),$
$K^{\ast}(892),$ $\omega(782),$ $\phi(1020)$\} with $J^{PC}=1^{--}$ are
listed. Being the second-lightest nonet, an enlargement of chiral perturbation theory that contains these states has been developed, see Ref. \cite{manohar}.
Moreover, the vector states belong to a homochiral multiplet \cite{pisarski}, hence the
strange-nonstrange mixing is expected to be small. In fact, one has:
\begin{equation}
\left(
\begin{array}
[c]{c}%
\omega(782)\\
\phi(1020)
\end{array}
\right)  =\left(
\begin{array}
[c]{cc}%
\cos\beta_{V} & \sin\beta_{V}\\
-\sin\beta_{V} & \cos\beta_{V}%
\end{array}
\right)  \left(
\begin{array}
[c]{c}%
\omega_{N}\\
\omega_{S}%
\end{array}
\right)  \text{ ,}%
\end{equation}
where the small isoscalar-vector mixing angle $\beta_{V}=-3.9^{\circ}$ implies
that $\omega(782)$ is mostly nonstrange and $\phi(1020)$ strange.

\item The chiral partners of vector mesons are the axial-vector mesons
\{$a_{1}(1260),$ $K_{1A}\equiv K_{1}(1270)$/$K_{1}(1400),$ $f_{1}(1285)$,
$f_{1}^{\prime}(1420)$\} with $J^{PC}=1^{++}$. The mass difference w.r.t. vector mesons is another clear manifestation of spontaneous chiral symmetry breaking. The isoscalar mixing angle is, as expected, also small, thus $f_{1}(1285)$ is predominately nonstrange and $f_{1}^{\prime
}(1420)$ strange. In the kaonic sector mixing with pseudovector mesons take
place, see below.

\item The nonet of pseudovector mesons $\{b_{1}(1235)_{,}K_{1B}\equiv
K_{1}(1270)/K_{1}(1400)$, $h_{1}(1170)$, $h_{1}(1415)\}$ with $J^{PC}=1^{+-}$
is also quite well known. The isoscalar states are again nonstrange and strange,
respectively. The kaonic states are not eigenstates of $C$, thus the state
$K_{1,A}$ belonging to the $J^{PC}=1^{++}$ axial-vector nonet and $K_{1,B}$
belonging to the $J^{PC}=1^{+-}$ pseudovector nonet mix:
\begin{equation}
\left(
\begin{array}
[c]{c}%
K_{1}(1270)\\
K_{1}(1400)
\end{array}
\right)  =\left(
\begin{array}
[c]{cc}%
\cos\varphi_{K} & -i\sin\varphi_{K}\\
-i\sin\varphi_{K} & \cos\varphi_{K}%
\end{array}
\right)  \left(
\begin{array}
[c]{c}%
K_{1,A}\\
K_{1,B}%
\end{array}
\right)  \text{ ,} \label{mixk1abfields}%
\end{equation}

\end{itemize}
with the large mixing angle $\varphi_{K}=\left(  56.4\pm4.3\right)  ^{\circ}$
\cite{divotgey,hatanaka}. On the other hand, the isoscalar mixing angle is not
yet known. It could be potentially large because the pseudovector mesons
belong to a heterochiral multiplet (just as pseudoscalar mesons), see also the discussions concerning pseudotensor mesons below.

\begin{itemize}
\item The chiral partners of the pseudovector mesons are the orbitally excited
vector mesons, the first ones in our table with $L=2$. They indeed fit quite
well with a regular nonet, but \textbf{one state is still not identified yet:
the mostly strange state }$\phi(???).$ In Ref. \cite{piotrowska} this state
was denoted with a putative  and yet undiscovered resonance
$\phi(1930).$ In the latest version of the PDG,
in the review of the quark model the assignment $\phi(2170)$ has been also
discussed, yet this interpretation does not fit with our results, both for
what concerns the mass and the decays of this state.

\item The next entry refers to the tensor states $\{a_{2}(1320)$,
$K_{2}^{\star}(1430),$ $f_{2}(1270),$ $f_{2}^{\prime}(1525)\}$ with
$J^{PC}=2^{++}$ \cite{burakovsky,bibrzycki,rodas} (for theoretical aspects see
\cite{annals}, yet for different interpretations Refs. \cite{geng,molina}). This
nonet, together with the pseudoscalar and vector mesons described above, is a
very well established nonet of quark-antiquark states that serve as an excellent
example for the validity of the quark model. The isoscalar states $f_{2}(1270)$
and $f_{2}^{\prime}(1525)$ are nonstrange and strange respectively, in
agreement with the homochiral nature of the underlying chiral multiplet.

\item The chiral partners of tensor mesons are the so-called axial axial-tensor
mesons $\{\rho_{2}(???)$, $K_{2}(1820)$, $\omega_{2}(???)$, $\phi_{2}(???)\}$
with $J^{PC}=2^{--}$, see also Refs. \cite{abreu,guo2}. 
\textbf{The natural question is: where are the states
}$\rho_{2},$\textbf{ }$\omega_{2},$\textbf{ and }$\phi_{2}$\textbf{? }It is
quite surprising that these states, that represent conventional mesons being
chiral partners of well known tensor states, could not be identified yet. In
Ref. \cite{axialtensors} they turn out to be quite wide, in agreement with
lattice \cite{dudek}, the main decay mode being the one into a vector-pseudoscalar mesonic pair.

\item Going further, one encounters the quite interesting pseudotensor mesons
$\{\pi_{2}(1670)$, $K_{2}(1770)$, $\eta_{2}(1645),$ $\eta_{2}(1870)\}$ with
$J^{PC}=2^{-+}$. These states fit rather well into the quark-antiquark picture,
under the assumption that the mixing angle in the isoscalar sector is large:
\begin{equation}
\left(
\begin{array}
[c]{c}%
\eta_{2}(1645)\\
\eta_{2}(1870)
\end{array}
\right)  =\left(
\begin{array}
[c]{cc}%
\cos\beta_{PT} & \sin\beta_{PT}\\
-\sin\beta_{PT} & \cos\beta_{PT}%
\end{array}
\right)  \left(
\begin{array}
[c]{c}%
\eta_{2,N}\equiv\sqrt{\frac{1}{2}}\,(\bar{u}u+\bar{d}d)\\
\eta_{2,S}\equiv\bar{s}s
\end{array}
\right)  \text{ ,}%
\end{equation}
with $\beta_{PT}\simeq-42^{\circ},$ similar to the case of pseudoscalar
mesons. The question here is: \textbf{is the mixing angle in the pseudotensor
really that large? If yes, is the chiral anomaly the reason for that? }This is
a relevant question because it would allow to investigate the nonperturbative
features of the chiral anomaly in a novel sector. Indeed, the pseudotensor
mesons belong to a heterochiral nonet, thus the mixing could be (potentially) large.

\item The chiral partners of the pseudotensor mesons are not listed in Table 1, but are expected to be heavier than 2 GeV. At present, they are completely unknown.

\item The last entry of Table 1 deals with the quite well established states
$\{\rho_{3}(1690)$, $K_{3}^{\star}(1780)$, $\omega_{3}(1670)$, $\phi
_{3}(1850)\}$ with $J^{PC}=3^{--}$. These states form also an almost ideal
nonet, whose isoscalar members $\omega_{3}(1670)$ and $\phi_{3}(1850)$ are
mostly nonstrange and strange, in agreement with the corresponding homochiral multiplet. The chiral partners of this nonet are also at present unknown. 
\end{itemize}

As next and final point, we have a quick look at some radially excited states with $n=2,$ see Table 2.

\begin{center}
Table 2: Some nonets of conventional radially excited mesons.%

\begin{tabular}
[c]{||l||l||l||l||l||l||l||}\hline\hline
$n^{2S+1}L_{J}$ & $J^{PC}$ &
\begin{tabular}
[c]{@{}l}%
$I=1$\\
$u\overline{d}$, $d\overline{u}$\\
$\frac{d\overline{d}-u\overline{u}}{\sqrt{2}}$%
\end{tabular}
&
\begin{tabular}
[c]{@{}l}%
$I=1/2$\\
$u\overline{s}$, $d\overline{s}$\\
$s\overline{d}$, $s\overline{u}$%
\end{tabular}
&
\begin{tabular}
[c]{@{}l}%
$I=0$\\
$\approx\frac{u\overline{u}+d\overline{d}}{\sqrt{2}}$%
\end{tabular}
&
\begin{tabular}
[c]{@{}l}%
$I=0$\\
$\approx s\overline{s}$%
\end{tabular}
& Refs.\\\hline\hline
$2^{1}S_{0}$ & $0^{-+}$ & $\pi(1300)$ & $K(1460)$ & $\eta(1295)$ &
$\eta(1440)$ & \cite{parga}\\\hline\hline
$2^{3}P_{0}$ & $0^{++}$ & $a_{0}(1950)$ & $K_{0}^{\star}(1950)$ &
$f_{0}(1790)$ & $f_{0}(2100)$ & \cite{parga}\\\hline\hline
$2^{3}S_{1}$ & $1^{--}$ & $\rho(1450)$ & $K^{\star}(1410)$ & $\omega(1420)$ &
$\phi(1680)$ & \cite{piotrowska}\\\hline\hline
$...$ & $...$ & $...$ & $...$ & $...$ & $...$ & \\\hline\hline
\end{tabular}

\bigskip
\end{center}

Some comments are in order:

\begin{itemize}
\item The pseudoscalar mesons are quite well known The states $\eta(1405)$ and $\eta(1475)$ can be actually identified with a single state $\eta(1440)$.

\item The excited scalar mesons are subject to large uncertainty, yet in Ref.
\cite{parga} an attempt toward their systematization is put forward.

\item The nonet of radially excited vector mesons is quite stable: masses and
decays fit well with the basic quark-antiquark assignment.

\item In general, it clear that, besides these examples, the radially excited states are still poorly known, leaving room for improvement.
\end{itemize}

\section{Conclusions}

In this work we have briefly reviewed the status of quark-antiquark states as resulting from hadronic flavor/chiral model(s) that we presented in a series
of papers listed in Tables 1 and 2. The main outcome is that the states listed
in these tables fit quite well with this basic $\bar{q}q$ picture, but some questions
(bold in the main text) are still open, in particular:

(i) Which is the $\phi$ meson of the orbitally excited vector meson multiplet?

(ii) Where are the axial-tensor mesons with $J^{PC}=2^{--}?$

(iii) Which is the value of the mixing angle in the isoscalar sector of
pseudotensor meson?

Finally, what about states that go beyond the $\bar{q}q?$ Besides the already
mentioned light scalar four-quark states and glueball(s), a special case is
the one of hybrid states.\ Quite recently, evidence toward a nonet of hybrid
($\bar{q}qg,$ where $g$ stands for a gluon) states with $J^{PC}=1^{-+}$ is emerging: besides the
well known $\pi_{1}(1600)$, the resonance $\eta_{1}(1855)$ has been newly
discovered \cite{beseta1}. Then, in\ Refs. \cite{hybrid1,hybrid2k} the
additional (not yet measured) states $\eta_{1}(1661)$ and $K_{1}(1761)$ have been discussed and their decays
have been evaluated, thus offering a prediction for future searches. 
In the near future, a better theoretical and experimental understanding of this
hybrid nonet is expected.

\section*{Acknowledgments}

I thank all my coworkers that contributed to the papers cited in Tables 1 and 2. Support from the Polish National
Science Centre (NCN) through the OPUS project 2019/33/B/ST2/00613 is acknowledged.

\end{document}